\documentclass[times,final]{elsarticle}

\usepackage{jcomp_af}
\usepackage{framed,multirow}

\usepackage{amssymb}
\usepackage{latexsym}
\usepackage{hyperref}
\usepackage{bigdelim}
\usepackage{tabularx}
\usepackage{booktabs}

\usepackage[singlelinecheck=false]{caption}

\usepackage{url}
\usepackage{xcolor}
\usepackage{algorithm}
\usepackage{algorithmicx}
\usepackage[fleqn]{amsmath}
\definecolor{newcolor}{rgb}{.8,.349,.1}

\DeclareMathAlphabet\mathbfcal{OMS}{cmsy}{b}{n}

\graphicspath{{../figs/}}

\journal{Manuscript for submission to the Journal of Computational Physics {(LLNL-JRNL-867233)}}

\begin{document}

\verso{J. R. Angus \textit{et al}}

\begin{frontmatter}

\title{Moment-preserving Monte-Carlo Coulomb collision method for particle codes}

\author{Justin Ray Angus\corref{cor1}}
\cortext[cor1]{Corresponding author}
\ead{angus1@llnl.gov}
\author{Yichen Fu}
\author{Vasily Geyko}
\author{Dave Grote}
\author{David Larson}

\address{Lawrence Livermore National Laboratory, Livermore, CA 94550, USA}


\begin{abstract}
Binary-pairing Monte-Carlo methods are widely used in particle-in-cell codes to capture effects of small angle Coulomb collisions. These methods preserve momentum and energy exactly when the simulation particles have equal weights. However, when the interacting particles are of varying weight, these physical conservation laws are only preserved on average. Here, we 1) extend these methods to weighted particles such that the scattering physics is correct on average, and 2) describe a new method for adjusting the particle velocities post scatter to restore exact conservation of momentum and energy. The efficacy of the model is illustrated with various test problems.
\end{abstract}


\end{frontmatter}


 
\section{Introduction}

Particle-in-cell (PIC) methods are a numerical approach for solving the phase-space continuity law known as the Vlasov equation, which governs collisionless kinetic processes in a plasma. The Vlasov equation is a subset of the more general Boltzmann equation that additionally includes short-range collisional processes. Binary-pairing Monte-Carlo collision (MCC) methods are a popular numerical approach for capturing collisional effects in a PIC code. Collisional processes in a fully ionized, ideal plasma are dominated by cumulative small angle Coulomb collisions as governed by the Landau-Fokker-Planck (LFP) equation \cite{bellan2008fundamentals}. 

The binary-pairing algorithms by Takizuka and Abe \cite{Takizuka1977205} (TA77) and Nanbu \cite{nanbu1997} (N97) are two standard MCC methods for modeling the LFP equation \cite{Wang20084308}. These algorithms are nearly identical, with the only difference being the formula used to determine the center-of-mass (CM) scattering angle for a pair of particles. For similar levels of accuracy, one can use a somewhat larger time step with N97 compared to TA77 \cite{Wang20084308}. Both TA77 and N97 identically conserve momentum and energy, consistent with the physical properties of the collision operator. However, these methods, as originally formulated, are limited to equally weighted particles. Extension of these methods to arbitrarily weighted particles has been considered by Miller \cite{miller1994}, Nanbu \cite{Nanbu1998639} and most recently by Higginson et al \cite{HIGGINSON2020109450}. In particular, it is shown in Ref.\,\cite{HIGGINSON2020109450} how the weighted-particle method by Nanbu in Ref.\,\cite{Nanbu1998639} produces erroneous results in certain scenarios, and an alternative formulation that produces the correct scattering physics on average for weighted particles is given. However, each of these weighted-particle methods only preserve momentum and energy on average after many scattering cycles.

Preservation of momentum and energy on average may work fine for homogeneous problems. However, for non-homogeneous problems, where particles transport to different regions on the grid with different physical properties, it may be difficult to understand the error introduced by non-exact conservation of momentum or energy during the application of the collision method. Furthermore, there is increasing interest in fully implicit PIC methods that are \textit{exactly} energy conserving \cite{CHEN20117018,MARKIDIS20117037}. Recently, these \textit{exactly} energy conserving methods have been coupled with models for collisions in Refs.\,\cite{angus2022,angus2023}. For these methods, it would not be ideal to throw away the \textit{exact} conservation of energy property when using weighted particles.

In this work, a binary-pairing MCC method for Coulomb collisions is presented that 1) produces the correct scattering physics for weighted particles on average and 2) maintains exact conservation of momentum and energy after each scattering cycle. The method has two parts. First, a generalization of the TA77 and N97 methods to arbitrarily weighted particles is derived starting from the general $N{\times}N$ method. The second part of the algorithm is a new method to adjust the velocity of the particles post-scatter such that exact momentum and energy conservation are restored. The method to restore energy conservation is based on inelastic scattering dynamics and works for both non-relativistic and relativistic particles.

The remainder of the paper is outlined as follows. Starting with a general discussion of the Boltzmann collision integral, the basics of small-angle Coulomb collisions in a plasma relevant to binary-pairing MCC methods is discussed in Sec.\,\ref{Sec:Boltzmann}. This is followed by a derivation of the weighted-particle extension of the TA77 and N97 methods in Sec.\,\ref{Sec:weightedMC}. The moment-correction method is described in Sec.\,\ref{Sec:MomentCorrection}. The efficacy of the method is illustrated with various test problems in Sec.\,\ref{Sec:Numerical_Tests}. The paper concludes with some further discussion and a summary in Sec.\,\ref{Sec:Discussion}.

\section{The Boltzmann collision integral} \label{Sec:Boltzmann}

The Boltzmann equation governing the phase-space distribution function, $f=f\left(\textbf{x},\textbf{v},t\right)$, for some arbitrary species is
\begin{equation}
\frac{\partial f}{\partial t} + \nabla_{\textbf{x}}\cdot\left(\textbf{v}f\right) + \nabla_{\textbf{v}}\cdot\left(\textbf{a}f\right) = \frac{\partial f}{\partial t}\bigg|_C, \label{Eq:Boltzmann}
\end{equation}
where $\textbf{x}$ and $\textbf{v}$ are the independent real-space and velocity-space coordinates, respectively, $\textbf{a}$ is the acceleration associated with long range, or collective, forces, and the term on the right of the equal sign is the Boltzmann collision integral. This term represents abrupt changes in velocity space owing to short-range interactions between particles. The left-hand side of Eq.\,\ref{Eq:Boltzmann} is the Vlasov equation, the phase-space continuity law for $f$. Particle-in-cell (PIC) methods are a numerical technique for solving this equation. Monte Carlo methods are commonly used in PIC codes to model the Boltzmann collision integral on the right-hand side (RHS), which is the topic of the present paper.

This Boltzmann collision integral can be written as a combination of sinks and sources in velocity space (see Appendix B of Ref.\,\cite{Lieberman_245}). For species $\alpha$ undergoing collisions with species $\beta$, this term can be written as
\begin{equation}
\frac{\partial f_{\alpha}}{\partial t}\bigg|_{C} = \frac{\partial f_{\alpha}}{\partial t}\bigg|_{\text{in}} - \frac{\partial f_{\alpha}}{\partial t}\bigg|_{\text{out}}, \label{Eq:CollisionIntegral}
\end{equation}
where
\begin{equation}
\frac{\partial f_{\alpha}}{\partial t}\bigg|_{\text{out}} \equiv f_{\alpha} \int\int f_{\beta} |\textbf{v}_{\alpha}-\textbf{v}_{\beta}|Id\Omega_{\alpha} d^3v_{\beta}, \ \text{and } \frac{\partial f_{\alpha}}{\partial t}\bigg|_{\text{in}} \equiv \int\int f'_{\alpha}f'_{\beta} |\textbf{v}'_{\alpha}-\textbf{v}'_{\beta}|I'd\Omega'_{\alpha} d^3v'_{\beta} \label{sink_and_source}
\end{equation}
are the velocity-space sink and source of particles, respectively. The quantity $I=I\left(|\textbf{v}_{\alpha}-\textbf{v}_{\beta}|,\theta_{\alpha}\right)$ here is the differential cross section that gives the probability that a particle from species $\alpha$ scatters into the solid angle $d\Omega_{\alpha}=d\phi_{\alpha}\sin\theta_{\alpha}d\theta_{\alpha}$. The total cross section is defined as $\sigma(|\textbf{v}_{\alpha}-\textbf{v}_{\beta}|) \equiv \int I d\Omega_{\alpha}$. The prime in the latter expression in Eq.\,\ref{sink_and_source} denotes post-scattered values. The sink term describes particles of species $\alpha$ leaving a specific region of velocity space by scattering \textit{into} all other locations in velocity space. The source term describes particles entering a specific region of velocity space by scattering \textit{out of} all other regions of velocity space.

In the analysis below, use is made of the Klimintovich representation of the species distribution functions:
\begin{equation}
f_{\alpha} = \sum_{i=1}^{N_{\alpha}}\frac{w_i}{\Delta V}\delta\left(\textbf{v}-\textbf{v}_i\right), \ \text{and} \ f_{\beta} = \sum_{j=1}^{N_{\beta}}\frac{w_j}{\Delta V}\delta\left(\textbf{v}-\textbf{v}_j\right), \label{Eq:Klimintovich}
\end{equation}
where $N_{\alpha}$ is the number of macro-particles (or just particles) used to describe species $\alpha$ in a given cell of volume $\Delta V$, $w_i$ is the weight of particle $i\in\alpha$, and similarly for species $\beta$. The spatial dependence of the distribution functions is neglected for simplicity. Furthermore, for notational simplicity, the subscript $i$ is used to refer to a specific particle belonging to species $\alpha$ and the subscript $j$ is used to refer to a specific particle belonging to species $\beta$. 

\subsection{Small angle Coulomb collisions} \label{Sec:Coulomb}

An expansion of the Boltzmann collision integral (Eqs.\,\ref{Eq:CollisionIntegral}-\ref{sink_and_source}) for small angle Coulomb collisions in an ideal plasma results in the LFP collision operator (see Ch.\,11 of \cite{schmidt1979physicsplasmas} and Ch.\,13 of Ref.\,\cite{bellan2008fundamentals}). This operator is of the drag-diffusion type with nonlinear drag and diffusion coefficients. For each particle pair, this process is a diffusive one for the CM scattering angle, $\Theta_{ij}$. The mean square CM scattering angle for a single collision of particle $i\in\alpha$ with particle $j\in\beta$ is computed by averaging $\Theta$ over all possible impact parameters (see Sec. 3.3 of Ref.\,\cite{Lieberman_245}) and can be expressed as
\begin{equation}
\left<\Theta_{ij}^2\right>\big|_{1} = \frac{2\pi b^2_{0,ij}}{\pi b^2_{\max}}\ln\Lambda.
\end{equation}
Here, $b_{\max}=\lambda_{De}$ is the maximum impact parameter for screened Coulomb collisions with $\lambda_{De}$ the plasma Debye length, $\ln\Lambda=\ln\left(2\lambda_{De}/b_{0,ij}\right)$ is the Coulomb logarithm, $b_{0,ij}/2 = q_{\alpha}q_{\beta}/\left(4\pi\epsilon_0\mu_{\alpha\beta}u_{ij}^2\right)$ is the minimum impact parameter with $\mu_{\alpha\beta}=m_{\alpha}m_{\beta}/\left(m_{\alpha}+m_{\beta}\right)$ the reduced mass, and $u_{ij}\equiv|\textbf{v}_{i}-\textbf{v}_j|$. The total number of collisions of particle $i$ with particle $j$ in time interval $\Delta t$ is $\pi b^2_{\max}n_ju_{ij}\Delta t$, where $n_j=w_j/\Delta V$ is the density associated with particle $j$. The total mean square CM angle for particle $i\in\alpha$ interacting with particle $j\in\beta$ during time interval $\Delta t$ is
\begin{equation}
\left<\Theta_{ij}^2\right>_{\text{all } b} = \left<\Theta_{ij}^2\right>\big|_{1}\pi b^2_{\max}\frac{w_j}{\Delta V}u_{ij}\Delta t = 2\pi b^2_{0,ij}\ln\Lambda \frac{w_j}{\Delta V}|\textbf{v}_{i}-\textbf{v}_j|\Delta t. \label{Eq:ThetaVar_coulomb}
\end{equation}
Being a diffusive process, the distribution of scattering angles $\Theta_{ij}$ after time $\Delta t$ is Gaussian with variance equal to that given in Eq.\,\ref{Eq:ThetaVar_coulomb}. The polar scattering angle $\Theta_{ij}$ is thus chosen by sampling from such a distribution. The total change of velocity of particle $i\in\alpha$ is obtained by sequentially performing this process for all $j\in\beta$. To see how this method connects with the LFP equation, the expected value for the change in velocity of particle $i\in\alpha$ is derived below.

The change in $\textbf{u}_{ij}$ parallel to itself after scattering through a small polar angle $\Theta_{ij}$ is $\Delta u_{||} = -2u_{ij}\sin^2\left(\Theta_{ij}/2\right)\approx -u_{ij}\Theta_{ij}^2/2$. After averaging $\Delta \textbf{u}_{ij}$ over the uniformly distributed azimuthal angle $\phi$, the parallel component is the only non-zero component \cite{schmidt1979physicsplasmas} and we can write
\begin{equation}
\left<\Delta\textbf{u}_{ij}\right>_{\text{all } b, \text{ all } \phi} = -\textbf{u}_{ij}\frac{1}{2}\left<\Theta^2_{ij}\right>_{\text{all }b}. \label{Eq:DeltaVij_allb_allphi}
\end{equation}
The change in velocity of particle $i\in\alpha$ in the lab frame is related to $\Delta\textbf{u}_{ij}$ by $m_{\alpha}\Delta\textbf{v}_i=\mu_{\alpha\beta}\Delta\textbf{u}_{ij}$. The final step in computing the expected value of $\Delta\textbf{v}_i$ in time interval $\Delta t$ is to sum Eq.\,\ref{Eq:DeltaVij_allb_allphi} for all $j\in\beta$:
\begin{equation}
\left<\Delta\textbf{v}_{i}\right> = -\frac{\mu_{\alpha\beta}}{m_{\alpha}}\sum_{j=1}^{N_{\beta}}\textbf{u}_{ij}\frac{1}{2}\left<\Theta^2_{ij}\right>_{\text{all }b} = -\frac{q^2_{\alpha}q^2_{\beta}\ln\Lambda}{4\pi\epsilon_0^2m_{\alpha}\mu_{\alpha\beta}}\sum_{j=1}^{N_{\beta}}\frac{w_j}{\Delta V}\frac{\textbf{u}_{ij}}{u^3_{ij}}\Delta t. \label{Eq:LDF_drag}
\end{equation}
After dividing by $\Delta t$, Eq.\,\ref{Eq:LDF_drag} is the well-known drag coefficient for the LFP equation (see Eq.\,13.28 of Ref.\,\cite{bellan2008fundamentals}). The tensor diffusion coefficient ($
\frac{1}{2}\left<\Delta v_i \Delta v_k\right>/\Delta t$), can additionally be obtained following a similar process \cite{schmidt1979physicsplasmas}. Note that each term in the sum in Eq.\,\ref{Eq:LDF_drag} scales linearly with the normalized scattering length $s_{ij}\equiv\frac{1}{2}\left<\Theta^2_{ij}\right>_{\text{all } b}$. The same is true for the diffusion coefficient.

\section{Monte Carlo method for weighted-particle Coulomb collisions} \label{Sec:weightedMC}

\begin{figure}[!ht] 
\centering
\includegraphics[scale=1.0]{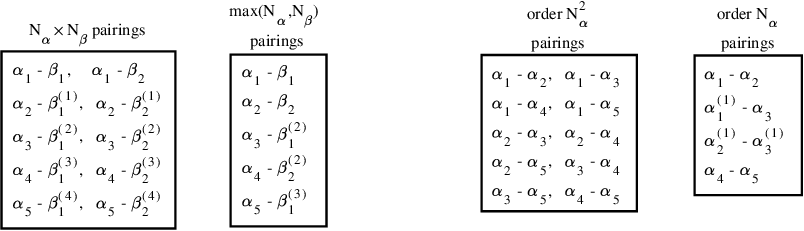}
\caption{Example pairings for order $N^2$ and order $N$ pairing methods. The left two panels are for inter-species collisions with $N_{\alpha}=5$ and $N_{\beta}=2$. The right two panels are for intra-species collisions with $N_{\alpha}=5$. The superscript in parentheses denote the number of times a particle has previously being paired. Note that when $N$ is odd, the first three particles for the order $N$ pairing method are done in an $N{\times}N$ way.} \label{TApairing}
\end{figure}

A method to extend binary-pairing algorithms for Coulomb collisions to weighted particles is described in this section. This is done for the $N{\times}N$ binary pairing method and the commonly used (and computationally practical) order $N$ binary pairing methods by Takizuka and Abe \cite{Takizuka1977205} and Nanbu \cite{nanbu1997}. For these latter methods, the scattering is done in such a way that the 
expected value of the change in velocity for each particle from each species is equal to that from the $N{\times}N$ method (i.e., Eq.\,\ref{Eq:LDF_drag}) on average after many pairings. This same constraint is used here to first extend the $N{\times}N$ scattering method to include weighted particles. Then, the reduction to an order $N$ binary-pairing method for weighted-particles follows exactly as it was first done in TA77. For reference, examples of the order $N{\times}N$ and order $N$ pairing methods considered in this work are illustrated in Fig.\,\ref{TApairing}. The order $N$ pairing method used here is the same as that described in TA77. 

\subsection{Order $N{\times}N$ binary pairing method}

The discussion in Sec.\,\ref{Sec:Coulomb} focused on the transport of a single particle $i\in\alpha$ interacting with species $\beta$. For numerical implementation, we seek to have the correct transport for each particle of each species. The normalized scattering length for particle $i\in\alpha$ interacting with particle $j\in\beta$ via cumulative small angle Coulomb collisions in an ideal plasma, and vice versa, are define as
\begin{equation}
s_{ij} \equiv \frac{1}{2}\left<\Theta_{ij}^2\right>_{\text{all }b} = \frac{q_{\alpha}^2q^2_{\beta}\ln\Lambda}{4\pi\epsilon_0^2\mu^2_{\alpha\beta}u_{ij}^3}\frac{w_j}{\Delta V}\Delta t, \ \text{and} \ s_{ji} \equiv \frac{1}{2}\left<\Theta_{ji}^2\right>_{\text{all }b} = \frac{q_{\alpha}^2q^2_{\beta}\ln\Lambda}{4\pi\epsilon_0^2\mu^2_{\alpha\beta}u_{ij}^3}\frac{w_i}{\Delta V}\Delta t. \label{Eq:sij_sji}
\end{equation}
If the particles are equally weighted ($w_i=w_j$ for all $i$ and $j$), then $s_{ij}=s_{ji}$ for all $i$ and $j$. One can then loop over all binary pairings (see Fig.\,\ref{TApairing}) and scatter each particle in the pair using the same common scattering angles. This approach obtains the correct scattering physics while also ensuring exact conservation of momentum and energy. However, when the interacting particles have different weights, then $s_{ij}\neq s_{ji}$. The reason is that macro-particles $i$ and $j$ are interacting with a different number of physical particles. One could use independent scattering angles for each particle in the pair. The scattering physics would be correct for each particle, but this approach requires sampling from two different Gaussian distribution functions for each binary pair.

We seek a weighted-particle implementation of the $N{\times}N$ method that only requires sampling from one Gaussian for each binary pair. This is achieved using the following normalized scattering length for the scattering pairs:
\begin{equation}
s\left(\alpha_i,\beta_j\right) = \frac{q_{\alpha}^2q^2_{\beta}\ln\Lambda}{4\pi\epsilon_0^2\mu^2_{\alpha\beta}u_{ij}^3}\frac{w_{\max}}{\Delta V}\Delta t, \ \ \ \left(\text{for order $N{\times}N$ pairings}\right) \label{sab_weighted}
\end{equation}
where $w_{\max}=\max(w_i,w_j)$ is used for the target density. Eq.\,\ref{sab_weighted} is the correct value as given in Eq.\,\ref{Eq:sij_sji} for the lower weight particle, but it is too large by $w_{\max}/\text{min}(w_i,w_j)$ for the larger weight particle. To compensate for this, the velocity vector for the larger weight particle is only updated with a probability $w_{\text{min}}/w_{\max}$. For simplicity, assume that $w_j>w_i$. Then, after many pairings between particle $i\in\alpha$ and $j\in\beta$, the average value of the normalized scattering length for the larger weight particle is
\begin{equation}
\left<s\left(\alpha_i,\beta_j\right)\frac{w_i}{w_j}\right> = \frac{q_{\alpha}^2q^2_{\beta}\ln\Lambda}{4\pi\epsilon_0^2\mu^2_{\alpha\beta}u_{ij}^3}\frac{w_i}{\Delta V}\Delta t = s_{ji},
\end{equation}
which is the correct value as given in Eq.\,\ref{Eq:sij_sji}. This approach is referred to as the rejection method and it is straight-forward to show that, in addition to achieving the correct scattering physics, momentum and energy are also preserved on average \cite{Nanbu1998639}.

\subsection{Order $N$ binary pairing method}

$N{\times}N$ methods are accurate, but the problem is that their computational cost scales with $N^2$. The first particle method for Coulomb collisions with computational cost of order $N$ is the seminal work by Takizuka and Abe, which uses $N_{\max}=\max\left(N_{\alpha},N_{\beta}\right)$ pairs per time step. The pairs are formed randomly and are such that each particle from the larger number population scatter exactly once per time step. The particles from the smaller number population scatter $N_{\max}/N_{\text{min}}$ times per time step on average. The normalized scattering length used in this method to the determine the scattering angle for each binary pair is computed using the value for the pair as used for the $N{\times}N$ method given by Eq.\,\ref{sab_weighted} multiplied by the number of particles for the population with the least number of particles: $N_{\text{min}}=\text{min}\left(N_{\alpha},N_{\beta}\right)$. This is expressed as
\begin{equation}
 s\left(\alpha_i,\beta_j\right) = \frac{q_{\alpha}^2q^2_{\beta}\ln\Lambda}{4\pi\epsilon_0^2\mu^2_{\alpha\beta}u_{ij}^3}\frac{w_{\max}N_{\text{min}}}{\Delta V}\Delta t, \ \ \ \left(\text{for order $N$ pairings}\right) \label{sab_TA77}
\end{equation}

To prove that this method gives the correct normalized scattering length on average for each particle we assume, without loss of generality, that $N_{\beta}>N_{\alpha}$. First, we consider equal weight particles as is considered in TA77. On average after many scatterings, the expected value of the change in velocity of particle $j\in\beta$ (the larger density population) is
\begin{equation}
\left<\Delta\textbf{v}_j\right> = -\frac{\mu_{\alpha\beta}}{m_{\beta}}\frac{1}{N_{\alpha}}\sum_{i=1}^{N_{\alpha}}\textbf{u}_{ij}s\left(\alpha_i,\beta_j\right) = -\frac{\mu_{\alpha\beta}}{m_{\beta}}\sum_{i=1}^{N_{\alpha}}\textbf{u}_{ij}s_{ji}, \label{sab_TA77_larger}
\end{equation} 
which is the correct value with $s_{ji}$ defined in the second expression in Eq.\,\ref{Eq:sij_sji}. The factor of $1/N_{\alpha}$ in the second expression is the probability that particle $j\in\beta$ is paired with particle $i\in\alpha$. The expected value of the change in velocity of particle $i\in\alpha$ (the smaller density population) is
\begin{equation}
\left<\Delta\textbf{v}_i\right> = -\frac{\mu_{\alpha\beta}}{m_{\alpha}}\frac{1}{N_{\beta}}\sum_{j=1}^{N_{\beta}}\textbf{u}_{ij}s\left(\alpha_i,\beta_j\right)\frac{N_{\beta}}{N_{\alpha}} = -\frac{\mu_{\alpha\beta}}{m_{\alpha}}\sum_{j=1}^{N_{\beta}}\textbf{u}_{ij}s_{ij}, \label{sab_TA77_smaller}
\end{equation} 
which is the correct value with $s_{ij}$ defined in the first expression of Eq.\,\ref{Eq:sij_sji}. The $N_{\beta}/N_{\alpha}$ factor in the second expression is present because each particle from species $\alpha$ is paired $N_{\beta}/N_{\alpha}$ times on average for each scattering event (see Fig.\,\ref{TApairing}). Thus, the scattering physics is correct for all particles on average when using $wN_{\min}/\Delta V$ for the target density. Since the particles are equally weighted here, momentum and energy are conserved exactly for each binary pair.

For weighted particles, Eq.\,\ref{sab_TA77} is used, but the velocity of the larger weight particle is only updated with probability $w_{\text{min}}/w_{\max}$, same as it is for the $N{\times}N$ method. The expected value of the change in velocity of particle $j\in\beta$ from the larger number population is
\begin{equation}
\left<\Delta\textbf{v}_j\right> = -\frac{\mu_{\alpha\beta}}{m_{\beta}}\frac{1}{N_{\alpha}}\sum_{i=1}^{N_{\alpha}}\textbf{u}_{ij}\frac{q_{\alpha}^2q^2_{\beta}\ln\Lambda}{4\pi\epsilon_0^2\mu^2_{\alpha\beta}u_{ij}^3}\frac{w_{\max}N_{\alpha}}{\Delta V}\Delta t P_{j}\left(\alpha_i,\beta_j\right) = -\frac{\mu_{\alpha\beta}}{m_{\beta}}\sum_{i=1}^{N_{\alpha}}\textbf{u}_{ij}s_{ji}. \label{sb_TA_weighted}
\end{equation}
For the last step, we made use of the fact that $P_{j}\left(\alpha_i,\beta_j\right)w_{\max} = w_{i}$ for all cases, where $P_{j}\left(\alpha_i,\beta_j\right)$ is the probability that particle $j\in\beta$ scatters when paired with particle $i\in\alpha$. The expected value of the change in velocity of particle $i\in\alpha$ from the smaller number population is
\begin{equation}
\left<\Delta\textbf{v}_i\right> = -\frac{\mu_{\alpha\beta}}{m_{\alpha}}\frac{1}{N_{\beta}}\sum_{j=1}^{N_{\beta}}\textbf{u}_{ij}\frac{q_{\alpha}^2q^2_{\beta}\ln\Lambda}{4\pi\epsilon_0^2\mu^2_{\alpha\beta}u_{ij}^3}\frac{w_{\max}N_{\alpha}}{\Delta V}\Delta tP_{j}\left(\alpha_i,\beta_j\right)\frac{N_{\beta}}{N_{\alpha}} = -\frac{\mu_{\alpha\beta}}{m_{\alpha}}\sum_{j=1}^{N_{\beta}}\textbf{u}_{ij}s_{ij}, \label{sa_TA77_weighted}
\end{equation} 
where again for the last step use is made of the fact that $P_{i}\left(\alpha_i,\beta_j\right)w_{\max} = w_{j}$ for all cases, where $P_{i}\left(\alpha_i,\beta_j\right)$ is the probability that particle $i\in\alpha$ scatters when paired with particle $j\in\beta$. 

In summary, the weighted-particle version of TA77 presented here is the same as the equal weight particle version with two changes. First, $w_{\max}$ is used for the density in the normalized scattering length for each binary pair. Second, the velocity of the larger weight particle is updated with probability $w_{\min}/w_{\max}$. The pairing method used here (see Fig.\,\ref{TApairing}), which is the same as in TA77, can also be used for the N97 method. The only difference between these two methods is the formula that takes the normalized scattering length and returns the scattering angle. These methods asymptote to the same variance as $\Delta t$ goes to zero \cite{HIGGINSON2020109450}.

Everything here is described for inter-species scattering. For intra-species scattering of a species with $N_{\alpha}$ particles in a cell, the pairing is done as described by TA77 and for weighted-particles one uses Eq.\,\ref{sab_TA77} with $N_{\text{min}} = N_{\alpha}-1$. When $N_{\alpha}$ is odd, the first three particles are paired in an $N{\times}N$ way and the normalized scattering length is reduced by a factor of two for these pairings to maintain the correct scattering length.

A slightly different method for extending the methods of TA77 and N97 to include weighted particles while preserving the scattering physics on average is described in Ref.\,\cite{HIGGINSON2020109450}. In that work, the normalized scattering length is computed using Eq.\,\ref{sab_TA77}, but with $N_{\min}$ replaced by $N_{\max}$, and $w_{\max}$ replaced by $w_{\max}/\mathcal{D}_{ij}$, where $\mathcal{D}_{ij}$ is duplicity factor for the particle pair defined by the number of times the particle from the lower number population is paired during a scattering event. Since the average duplicity factor per binary pair is $\left<\mathcal{D}_{ij}\right>=N_{\max}/N_{\min}$, the method described in this work is like that from Ref.\,\cite{HIGGINSON2020109450} when using the average duplicity factor for each particle pair.

\section{Moment correction method} \label{Sec:MomentCorrection}

The Monte Carlo method for weighted-particle Coulomb collisions described in the previous section only preserves momentum and energy on average after many pairings. Several methods have been used before to restore exact momentum and/or energy conservation after application of non-conservative scattering algorithm, such as the models by Jones \cite{JONES1996169} and Lemons \cite{Lemons2009}. In the method by Jones, the post-scattered velocity of each particle is linearly transformed by shifting and rescaling the post-scattered velocity of each particle The method by Lemons is similar, but the linear transformation is applied to the change in velocity from scattering rather than to the velocity itself. In these methods, each particle is shifted and rescaled by the same common values, which are readily computed using the laws of conservation of momentum and energy for non-relativistic particles. For relativistic particles, the shift parameter is still readily computed, but a different formula for rescaling the velocities needs to be used \cite{COHEN201333}. The formula used in Ref.\,\cite{COHEN201333} is for like-species collisions and requires the energy-correction factor to be of the form $1+\alpha$ with $|\alpha|\ll 1$. The parameter $\alpha$ is obtained by expanding the kinetic energy to first order in $\alpha$ and is therefore not exact. A weighted-particle scattering method is used by Sentoko and Kemp in Ref.\,\cite{SENTOKU20086846} that preserves energy exactly and works for relativistic particles, but momentum is only preserved on average.

Here, a new method is proposed for restoring momentum and energy conservation. Momentum is restored by applying a linear shift to the particles velocities like that described above. The method to restore energy conservation is based on kinematics for zero-angle inelastic scattering of binary pairs. Being based on scattering kinematics, this method works for both non-relativistic and relativistic particles. The method works as follows. First, the weighted-particle scattering algorithm is applied to two interacting species. The particle velocity before and post-scatter are denoted, respectively, as, $\textbf{v}_p^{bs}$ and $\textbf{v}_p^{ps}$. Second, momentum conservation is restored by shifting the particle velocities as
\begin{equation}
\textbf{v}_p^{pm} = \textbf{v}_p^{ps}-\textbf{B}w_p \ \Rightarrow \  \textbf{B} = \frac{\sum_pm_p\left(\textbf{v}^{ps}_p-\textbf{v}_p^{bs}\right)}{\sum_{p}m_pw_p}, \label{Eq:vp_pm}
\end{equation}
where $m_p=m_sw_p$ is the mass of particle $p\in s$ and $\textbf{v}_p^{pm}$ is the post-momentum corrected velocity of particle $p$. For relativistic particles, one shifts the proper velocity $\textbf{u}_p=\gamma_p\textbf{v}_p$ with $\gamma_p=1/\sqrt{1-|\textbf{v}_p|^2/c^2}$. The difference between the momentum correction used here and that used in Refs.\,\cite{JONES1996169,Lemons2009} is that here the shift is biased to the larger weighted particles. After momentum conservation has been restored, the total violation in energy conservation in a cell is computed as
\begin{equation}
\Delta E = \sum_{p}m_p\frac{1}{2}|\textbf{v}^{pm}_p|^2 - KE^{bs}, \label{Econs}
\end{equation}
where $KE^{bs}=\sum_pm_p|\textbf{v}_p^{bs}|^2/2$ is the total kinetic energy in a cell before scattering. The final step is to adjust the energy of the particles such that $\Delta E$ is re-absorbed. As mentioned above, the method we use is based on zero-angle inelastic scattering kinematics of binary pairs. It is instructive to first go over this process. It is done here for non-relativistic particles. An extension to relativistic particles is given in Appendix A. 

The post-collision velocity of two particles undergoing scattering in the CM frame can be written as
\begin{eqnarray}
\textbf{v}'_1 &=& \textbf{v}_1 + \frac{\mu_R}{m_1}\left(\textbf{v}'_R-\textbf{v}_R\right), \label{vp1_inelastic} \\
\textbf{v}'_2 &=& \textbf{v}_2 - \frac{\mu_R}{m_2}\left(\textbf{v}'_R-\textbf{v}_R\right), \label{vp2_inelastic}
\end{eqnarray}
where $\mu_R=m_1m_2/(m_1+m_2)$ is the reduced mass, $\textbf{v}_R=\textbf{v}_1-\textbf{v}_2$ is the relative velocity, and the prime superscript denotes post scatter. The energy conservation law for an inelastic event with change in energy $U$ is
\begin{equation}
\frac{1}{2}\mu_R|\textbf{v}'_R|^2 = \frac{1}{2}\mu_R|\textbf{v}_R|^2 - U. \label{consE_inelastic}
\end{equation}
For zero-angle scattering, the post-scatter relative velocity vector is obtained directly from Eq.\,\ref{consE_inelastic} and is given as
\begin{equation}
\textbf{v}'_R = \textbf{v}_R\sqrt{1 - \frac{U}{0.5\mu_R|\textbf{v}_R|^2}}. \label{vrel_inelastic}
\end{equation}

The weighted-scattering methods described in Sec.\,\ref{Sec:weightedMC} are such that the correct scattering physics is preserved on average. When adjusting the distribution function post-scatter to restore momentum and energy conservation, one should minimize how much the distribution function is perturbed to maintain this property. Here, the energy is corrected by forming binary pairs and adjusting the CM energy using Eqs.\,\ref{vp1_inelastic}-\ref{vp2_inelastic}. To try and minimize the perturbation, the potential $U$ used for a given binary pair is chosen to be some small percentage of the CM energy of the binary pair, that is $U=\text{sign}(\Delta E)f_E0.5\mu_R|\textbf{v}_p|^2$ where $f_E$ is a user-specified small parameter referred to as the energy-correction factor. If $\Delta E<0$, then energy needs to be added back to the particles to restore energy conservation and vice versa. The binary pairs are formed randomly. After each binary pair, $\Delta E$ is adjust accordingly and the loop proceeds until $\Delta E=0$. To restore energy conservation \textit{exactly}, $|U|$ is chosen as:
\begin{equation}
U = \text{sign}\left(\Delta E\right)\min\left(|\Delta E|,f_E0.5\mu_R|\textbf{v}_p|^2\right). \label{U_inelastic}
\end{equation}

The free parameter $f_E$ in this model is the percentage that one allows the CM energy of the binary pairs to change. If $f_E$ is too large, then the distribution function may be perturbed too much, and the scattering physics can be disrupted. However, more binary pairs are needed to restore energy conservation for smaller values of $f_E$, which means the computational cost of the algorithm increases. Typical values of $f_E$ used here are between $0.02$ and $0.05$.

Another important things to note about the application of Eqs.\,\ref{vp1_inelastic}, \ref{vp2_inelastic}, and \ref{vrel_inelastic} used in this context is that the masses used are the weighted particle ones ($m_1=m_sw_1$ and $m_2=m_sw_2$). For this reason, it can be useful to sort the particles by weight before forming the binary pairs such that the particles are more likely to be paired with a similar weight particle. Furthermore, this ensures that the energy-correction is biased to the larger weight particles. This is preferable because the error in momentum and energy conservation associated with each application of the weighted-scattering algorithm reside in the larger weight particles, which only scatter probabilistically. Another way to say this is that it is preferable to minimize adjusting the velocity of the lower weight particles, which are typically scattered correctly at each application of the weighted-scattering algorithm.

The full weighted-scattering algorithm including steps to restore momentum and energy conservation is outlined in Algorithm\,\ref{WeightedScattering}. This algorithm works for both order $N$ TA-like pairings and for the $N{\times}N$ method. The only difference is in the formula used for the normalized scattering length for the binary pairs. Algorithm\,\ref{WeightedScattering} is also written to be valid for intra-species and inter-species collisions. For inter-species scattering, one needs to choose how to partition the net violation in energy conservation between the two species. In this work, $\Delta E$ is partitioned as follows
\begin{equation}
\Delta E_s = \frac{\bar{w}_sE_s^{pm}}{\bar{w}_{\alpha}E_{\alpha}^{pm}+\bar{w}_{\beta}E_{\beta}^{pm}}\Delta E, \ \text{for} \ s=\alpha,\beta. \label{Eq:DeltaEs}
\end{equation}
In Eq.\,\ref{Eq:DeltaEs}, $\bar{w}_s = \sum_{p\in s}w_p/N_{s}$ is the average particle weight of species $s$ in the cell and $E_s^{pm}$ is the total energy in the cell of species $s$ post momentum correction. The weighting of $\Delta E_s$ with $\bar{w}_s$ is for the purpose of biasing the energy correction to larger weight particles.

\begin{algorithm}
\caption{Moment-preserving weighted Coulomb collision method.}
\begin{algorithmic}[1]
\item
Compute the initial energy and momentum of each species in the cell.\\
Formulate random binary pairs as described in Fig.\,\ref{TApairing}.
\Statex Loop over pairs and for each pair:
\State\hspace{\algorithmicindent} Compute the scattering angle using $s(\alpha_i,\beta_j)$ given in Eq.\,\ref{sab_TA77} for TA-like or Eq.\,\ref{sab_weighted} for $N{\times}N$.
\State\hspace{\algorithmicindent} Use the formula from TA77 or N97 to compute the CM scattering angle from $s(\alpha_i,\beta_j)$
\State\hspace{\algorithmicindent} Update $\textbf{v}_p$ for the smaller-weight particle of the pair.
\State\hspace{\algorithmicindent} Update $\textbf{v}_p$ for the larger-weight particle of the pair with probability $w_{\min}/w_{\max}$.\\
Restore momentum conservation by shifting the velocity of each particle using Eq.\,\ref{Eq:vp_pm}.\\
Compute the net violation in energy conservation $\Delta E$ using Eq.\,\ref{Econs}.\\
For inter-species scattering, partition $\Delta E$ using Eq.\,\ref{Eq:DeltaEs}.\\
(Optional step) Sort particles of each species in descending order by weight.
\Statex For each species $s$, loop over particles forming binary pairs and for each pair:
\State\hspace{\algorithmicindent} Compute $\textbf{v}'_R$ in Eq.\,\ref{vrel_inelastic} using $U$ from Eq.\,\ref{U_inelastic}.
\State\hspace{\algorithmicindent} Modify $\textbf{v}_p$ of each particle using Eqs.\,\ref{vp1_inelastic}-\ref{vp2_inelastic}.
\State\hspace{\algorithmicindent} Adjust net energy violation: $\Delta E_s \mathrel{-}=U$.
\State\hspace{\algorithmicindent} Break loop when $\Delta E_s =0$.
\end{algorithmic}  \label{WeightedScattering}
\end{algorithm}

\section{Numerical tests} \label{Sec:Numerical_Tests}

Results from several numerical tests using the moment-preserving weighted-particle Monte-Carlo scattering algorithm are presented in this section. The first test considered is Test 1 from Ref.\,\cite{HIGGINSON2020109450}, which considers two populations within the same species relaxing to a common velocity and temperature. Same as in Ref.\,\cite{HIGGINSON2020109450}, simulations of this test are performed using a variety of different weight ratios for the two populations to rigorously verify the accuracy of the weighted-particle scattering method. The second test is Test 2 from Ref.\,\cite{HIGGINSON2020109450}, which is like Test 1, but includes three populations belonging to two different species to test inter-species scattering. The third test is an electron-ion thermalization test for a fully ionized carbon plasma. Unless stated otherwise, the particles are sorted in descending order of particle weights prior to adjusting the velocities to restore energy conservation.

\subsection{Test 1}

Consider two populations, call them $A$ and $B$, of fully ionized carbon 12 atoms ($A = 12$, $Z = 6$) with $m_i = Am_u-Zm_e$, where $m_u$ is the atomic mass unit and $m_e$ is the electron mass. The initial density, temperature, and velocity of population $A$ are $n_A = 10^{19}/\text{cm}^3$, $T_A = 500$\,eV, and $U_A=+655$\,km/s. Those for population $B$ are $n_B = 10^{20}/\text{cm}^3$, $T_B = 500$\,eV, and $U_A=0$\,km/s. Momentum conservation gives $U_f=U_An_A/\left(n_A+n_B\right)=59.55$ for the equilibrium value for the velocity of each species. Energy conservation gives $T_f=1.97$\,keV for each species. The simulations are performed in quasi 0D using a 1D grid with 180 grid cells. The number of particles in Table\,\ref{test1_parameters} below summarizing the numerical setup for this test refer to the number of particles per cell. A fixed value of $\Lambda=10$ is used for the Coulomb logarithm. The particle positions are not advances and there are no fields involved. The simulations are velocity-space only and the results presented below are averaged values on the grid.

\begin{table*}[ht]
\centering
\captionsetup{font={normalsize,singlespacing}}
\captionsetup{skip=4pt}
\captionsetup{justification=centering}
\centering
\caption{Test 1 simulation parameters.}
\begin{tabular*}{9cm}{@{\extracolsep{\fill}} l c ccc ccc}
\toprule
 && $N_A$ & $N_B$ & $N_A{:}N_B$  & $w_A{:}w_B$ & $\Delta t$\,[fs] & $f_E$ \\
\midrule
T1a  &&  400 & 4,000 & 1:10 & 1:1 & 50 & 0.05 \\
T1b  &&  400 & 400 & 1:1 & 1:10 & 50 & 0.05 \\
T1c  &&  4,000 & 400 & 10:1 & 1:100 & 15  & 0.02 \\
T1d  &&  200 & 8,000 & 1:40 & 4:1 & 50  & 0.05 \\
\bottomrule
\end{tabular*} \label{test1_parameters}
\end{table*}

\begin{figure}[!ht] 
\centering
\includegraphics[scale=1.0]{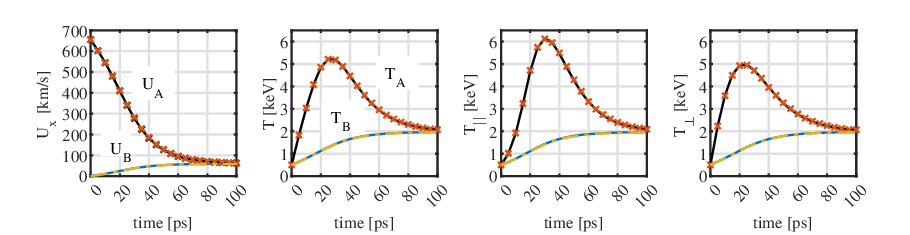}
\caption{Mean velocity and temperature profiles for populations $A$ and $B$ for tests T1a and T1b. The solid black and solid blue curves are for populations $A$ and $B$, respectively, from test T1a (equally weighted particles). The red $\times$ markers and the dashed yellow curves are from test T1b, where the particle weights of the denser population B are 10$\times$ larger than that for the lower density population A.} \label{T1a_T1b}
\end{figure}

The numerical parameters for tests T1a-T1d are summarized in Table.\,\ref{test1_parameters}. One thing to note is that the number of particles per cell used here is 10 times less than that considered in Ref.\,\cite{HIGGINSON2020109450}. Results for tests T1a (equal particle weights) and tests T1b (equal number of particles) are shown in Fig.\,\ref{T1a_T1b}. The results from test T1a are considered the baseline solution. Results from test T1b shown in this figure, where the ratio of particles weights for the low-density population to the high-density populations is 1:10, are indistinguishable from test T1a. The relaxation of the mean velocity and temperature profiles shown in Fig.\,\ref{T1a_T1b} can additionally be compared with the same results presented in Fig.\,3 of Ref.\,\cite{HIGGINSON2020109450}.

\begin{figure}[!ht] 
\centering
\includegraphics[scale=1.0]{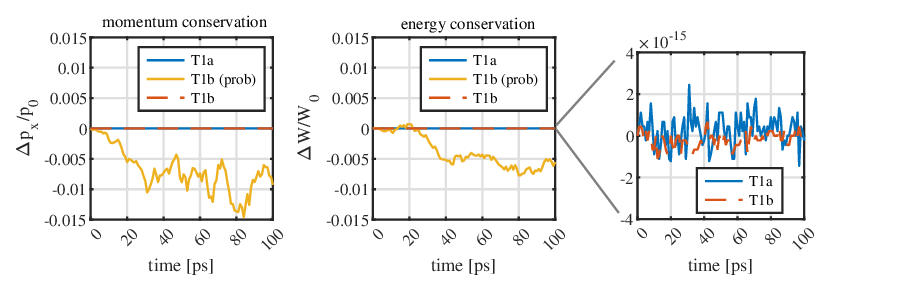}
\caption{Conservation of x-momentum (left panel) and energy (middle and right panels) for test T1b with the moment correction method applied (dashed red curves) and without (yellow curves). Results are compared with those from test T1a, which uses equal particle weights and thus naturally conserves momentum and energy. The right panel shows the relative change in total energy for tests T1a and T1b using the moment correction method on a scale where it is evident that the violation in energy conservation is at machine precision.} \label{Test1_momentConservations}
\end{figure}

The violation in total x-momentum and energy for test T1b with and without the moment-correction algorithm applied are shown in Fig.\,\ref{Test1_momentConservations}. Although not shown, the velocity and temperature relaxation results shown in Fig.\,\ref{T1a_T1b} for test T1b are similar independent of using the moment-correction algorithm. The relative error in momentum and energy conservation is between $0.5-1.5$\% when not correcting the moments and using unequal weight particles. Momentum and energy conservation are preserved to machine precision with weighted particles when applying the moment-correction method, as seen in the right panel of Fig.\,\ref{Test1_momentConservations}.

\begin{figure}[!ht] 
\centering
\includegraphics[scale=1.0]{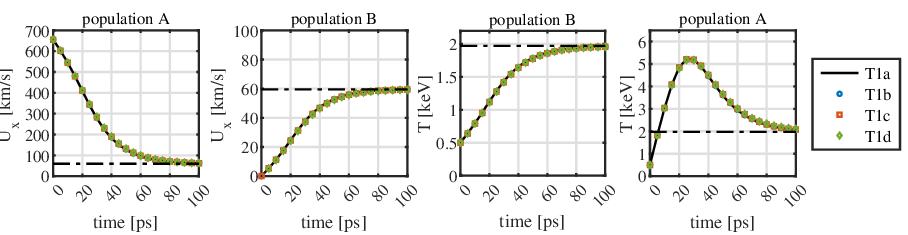}
\caption{Mean velocity and temperature profiles of populations A and B for tests T1a - T1d. See Table\,\ref{test1_parameters} for simulation parameters. The dashed-dotted black curves are the equilibrium solutions.} \label{T1abcd}
\end{figure}

Results from all four test cases described in Table.\,\ref{test1_parameters} are shown in Fig.\,\ref{T1abcd}. The relaxation of the velocity and temperature profiles for both species match well for all cases. These results show 1) that the weighted scattering algorithm achieves the correct scattering physics for a wide range of weighted-particle scenarios, and 2) the moment-correction method can be applied without significantly altering the scattering physics. It is worth noting here that the original method for doing weighted-particle Coulomb scattering by Nanbu in Ref.\,\cite{Nanbu1998639} does not achieve the correct relaxation for any of the unequal weight particles tests for this problem (see Fig.\,6 of Ref.\,\cite{HIGGINSON2020109450}).

\begin{figure}[!ht] 
\centering
\includegraphics[scale=1.0]{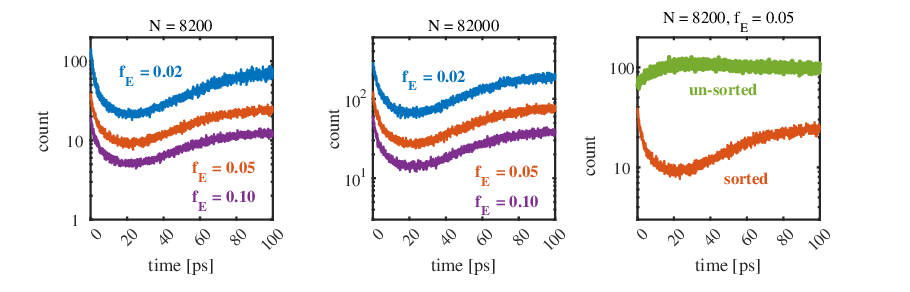}
\caption{Time profile of the number of binary pairs needed to restore energy conservation for test T1d. The left and middle panels show results for various values of the energy-correction factor $f_E$. The results in the middle panel are obtained using $10\times$ more particles than used for the left panel. The right panel shows how these results differ for this test if one chooses not to sort the particles by weight prior to applying the energy correction.} \label{T1d_count}
\end{figure}

\begin{figure}[!ht] 
\centering
\includegraphics[scale=1.0]{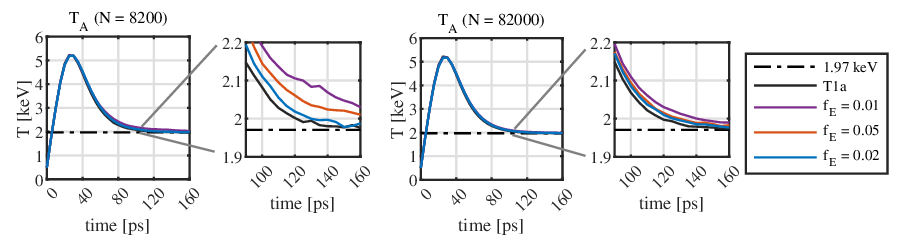}
\caption{Long-time profiles for the temperature of the low-density population A from T1d simulations with various values of the energy-correction factor. The results in the right two panels are from simulations using $10\times$ the number of particles compared to those for the left two panels.} \label{T1d_fEscan}
\end{figure}

Time profiles for the number of binary pairs required to restore energy conservation for test T1d are shown in Fig.\,\ref{T1d_count} for various values of the energy-correction factor $f_E$. Results are shown in the middle panel of this figure when using $10\times$ more particles. The number of binary pairs needed to restore energy conservation scales approximately linear with $f_E$ and approximately as $\sqrt{N}$. The effects of $f_E$ and $N$ on $T_A$ from test T1d are shown in Fig.\,\ref{T1d_fEscan}. The zoomed in inserts in this figure show the solution converging to the equally weighted particle solution as $f_E\rightarrow 0$ and with increasing $N$.

\subsection{Test 2}

The second test considered is similar to Test 1, but populations A and B are now separated into different species, species $\alpha$ and species $\beta$, and a third population C is added to species $\beta$. Population C is a low-density population with the following initial conditions: $n_C = 10^{18}/\text{cm}^3$, $T_C = 500$\,eV, and $U_A=-655$\,km/s. The simulation parameters for Test 2 are summarized in Table\,\ref{test2_parameters}. A variety of different weight combinations for the different populations are considered to rigorously verify the weighted-particle scattering algorithm. Tests T2a-T2e are the same as those considered in Ref.\,\cite{HIGGINSON2020109450}, but with $10\times$ less particles, same as for Test 1. Sorting is used for $\alpha-\beta$ and $\beta-\beta$ scattering but is not used for $\alpha-\alpha$ scattering since each of the particles in species $\alpha$ have equal weights.

\begin{table*}[ht]
\centering
\captionsetup{font={normalsize,singlespacing}}
\captionsetup{skip=4pt}
\captionsetup{justification=centering}
\centering
\caption{Test 2 simulation parameters.}
\begin{tabular*}{12cm}{@{\extracolsep{\fill}} l c cccc cccc}
\toprule
 && $N_A$ & $N_B$ & $N_C$ & $N_A{:}N_B{:}N_C$ & $N_{\alpha}{:}N_{\beta}$ & $w_A{:}w_B{:}w_c$ & $\Delta t$\,[fs] & $f_E$ \\
\midrule
T2a  &&  400 & 4,000 & 200 & 2:20:1 & 1:10.5 & 1:1:1 & 50 & 0.05 \\
T2b  &&  400 & 4,000 & 800 & 1:10:2 & 1:12 & 4:4:1 & 50 & 0.05 \\
T2c  &&  400 & 800 & 4,000 & 1:2:10 & 1:12 & 20:100:1 & 50 & 0.05 \\
T2d  &&  4,000 & 500 & 100 & 40:5:1 & 40:6 & 1:80:20 & 50 & 0.05 \\
T2e  &&  4,000 & 100 & 500 & 40:1:5 & 40:6 & 1:400:4 & 50 & 0.05 \\
\bottomrule
\end{tabular*} \label{test2_parameters}
\end{table*}

\begin{figure}[!ht] 
\centering
\includegraphics[scale=1.0]{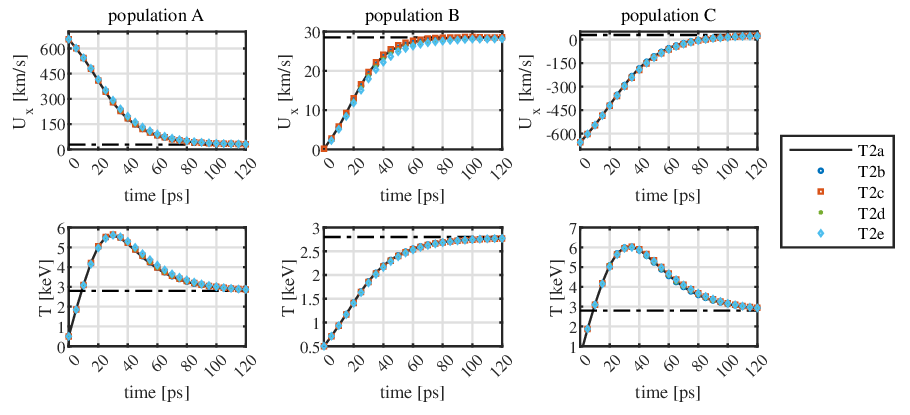}
\caption{Mean velocity and temperature profiles for populations A, B, and C for tests T2a - T1e. See Table\,\ref{test2_parameters} for simulation parameters. The dashed-dotted black curves are the equilibrium solutions.} \label{T2abcde}
\end{figure}

Time-profiles of the mean velocity and temperature of populations A, B, and C for tests T2a-T2e are shown in Fig.\,\ref{T2abcde}. Each of the quantities for each of the populations shown in Fig.\,\ref{T2abcde} are observed to relax to the steady state solutions. Furthermore, the relaxation to the steady state values is observed to agree well for all particle weighting combinations considered. There is a small, but noticeable deviation of the $U_x$ profile for population $B$ for test T2e, which is an extreme case where $w_A{:}w_B{:}w_C = 1{:}400{:}4$. Furthermore, there are only 100 particles per cell for population $B$ in this test. This small discrepancy goes away with increasing the number of particles per cell.

\begin{figure}[!ht] 
\centering
\includegraphics[scale=1.0]{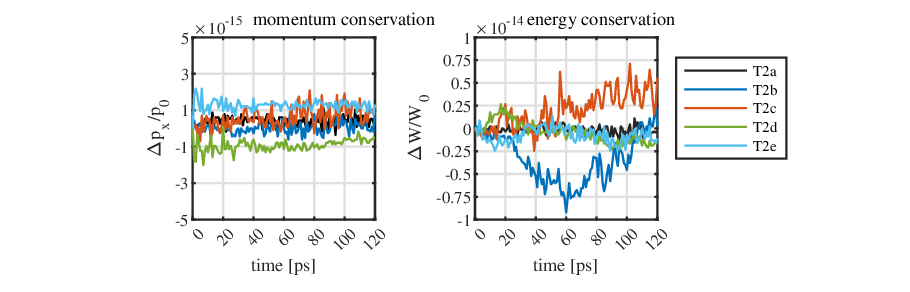}
\caption{Conservation of x-momentum (left panel) and energy (middle and right panels) for test T2a-T2e.} \label{Test2_momentConservations}
\end{figure}

The ability to maintain rigorous conservation of momentum and energy for Tests T2a-T2e is shown in Fig.\,\ref{Test2_momentConservations}. The relative change in total x-momentum and energy is shown in this figure to be on the order of machine-precision for each of these tests.

\subsection{Test 3}

For the final test, electron-ion thermalization of a fully ionized carbon plasma is considered ($A = 12$, $Z = 6$). The electron and ion distributions are initialized as non-drifting Maxwellians with the following conditions: $n_i = 1{\times}10^{23}/$cm$^3$, $T_i = 50$\,eV, $n_e = 6{\times}10^{23}/$cm$^3$, and $T_e = 150$\,eV. The equilibrium solution for the temperature is $T_f = 136$\,eV. The simulations are performed using a 2D periodic domain on a $40{\times}40$ grid with uniform spacing equal to $1.3285$\,nm. For all simulations, $\Delta t = 3.54{\times}10^{-4}$\,fs is used for the time step and $\ln\Lambda = 3$ is used as the Coulomb logarithm. This time step is roughly $100$ times smaller than the characteristic electron collision time: $\tau_e[\text{s}] = 3.44{\times}10^{5}T_e^{3/2}[\text{eV}]/\left(Zn_e[1/\text{cm}^3]\Lambda\right)$.

\begin{figure}[!ht] 
\centering
\includegraphics[scale=1.0]{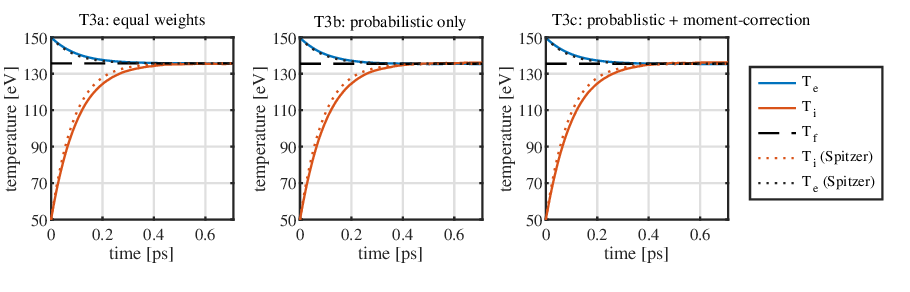}
\caption{Species temperature relaxation results for a fully ionized carbon plasma. In each figure, the dashed black curve is the equilibrium solution, and the dotted curves are obtained from a 0D Spitzer thermalization model as described in the text. Equally weighted particles are used for the results shown in the left panel. The results shown in the middle and right panels are from simulations that use an equal number of simulation particles for both species, without (middle panel) and with (right panel) the moment correction method applied.} \label{Test3}
\end{figure}

Simulation results for this test problem are shown in Fig.\,\ref{Test3}. Here, time profiles for the average electron and ion temperatures on the 2D grid are shown for three cases. The number of ion particles per cell is set to $N_i=512$ for each of these simulations. The results in the left panel are obtained using equally weighted particles (T3a), in which case the number of electron particles per cell is $N_e=6N_i=3072$. The simulations corresponding to the middle and right panels use $N_e=N_i=512$, in which chase $w_e=6w_i$. The results in the middle and right panels are obtained without (T3b) and with (T3c) the moment-correction algorithm, respectively. The electron and ion temperature profiles from each of these simulations are observed to match well. Furthermore, the relaxation of $T_e$ and $T_i$ shown in Fig.\,\ref{Test3} are also observed to agree well with results obtained from a 0D Spitzer model:
\begin{equation}
\frac{dT_e}{dt} = -\nu_T\left(T_e-T_i\right); \ \frac{dT_i}{dt} = -Z\frac{dT_e}{dt}, \label{0DSpitzer}
\end{equation}
where $\nu_T = 2m_e/M_i/\tau_e$ is the thermalization rate.

\begin{figure}[!ht] 
\centering
\includegraphics[scale=1.0]{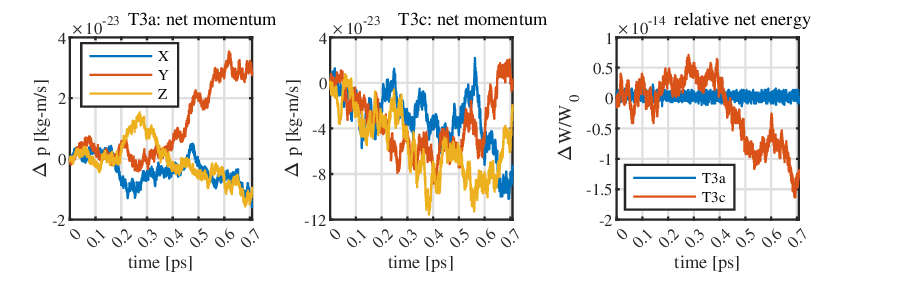}
\caption{Change in total momentum (left and middle panels) and relative change in total energy (right panel) correspondig to test T3a and T3c shown in Fig.\,\ref{Test3}.} \label{Test3_momentConservations}
\end{figure}

The ability to use weighted particles while conserving momentum and energy for test T3c is shown in Fig.\,\ref{Test3_momentConservations}. The left two panels in this figure show the change in total momentum in each direction. The amplitude of the change in momentum for test T3c shown in the middle panel is about 2-3 times larger than that for the simulation with equal weights shown in the left panel. In both cases, the change in momentum is close to machine precision and behaving like a random walk. Similarly, the relative change in energy for test T3c also behaves like a random walk, but with a step size that is about $10{\times}$ larger than for test T1a where equal weight particles are used. The reason for this is unclear. It may be due to the large electron-ion mass ratio. In either case, energy is conserved nearly to machine-precision for weighted-particles with the moment correction algorithm applied.

\section{Discussion and summary} \label{Sec:Discussion}

\subsection{Analysis of perturbations to distribution functions}

It is worthwhile to analyze how the energy-correction scheme described above perturbs the species distribution functions. For simplicity, assume $\Delta E$ after momentum is restored is negative. Using Eq.\,\ref{U_inelastic} in Eq.\,\ref{vrel_inelastic}, we can write
\begin{equation}
\textbf{v}'_R - \textbf{v}_R = \textbf{v}_R\sqrt{1+f_E} - \textbf{v}_R \approx \frac{f_E}{2}\textbf{v}_R. \label{deltaVrel}
\end{equation}
It is additionally assumed that the particles are sorted by weight such that $m_1\approx m_2$. Then, using Eq.\,\ref{deltaVrel}, the post-energy-correction particle velocities in Eqs.\,\ref{vp1_inelastic}-\ref{vp2_inelastic} become
\begin{eqnarray}
\textbf{v}'_1 &=& \textbf{v}_1 + \frac{f_E}{4}\left(\textbf{v}_1-\textbf{v}_2\right), \label{vp1_inelastic_2} \\
\textbf{v}'_2 &=& \textbf{v}_2 - \frac{f_E}{4}\left(\textbf{v}_1-\textbf{v}_2\right). \label{vp2_inelastic_2}
\end{eqnarray}
We analyze Eqs.\,\ref{vp1_inelastic_2}-\ref{vp2_inelastic_2} in several different scenarios. First, assume the velocity of the two particles have similar magnitudes, but are in opposite directions. In this case, the velocity of both particles is scaled by $1+f_E/2$. If the velocity of the two particles is similar in magnitude and are in the same direction, then there is essential no modification to the particle velocities because the CM energy is approximately zero. Another case to consider is where the particle speeds are much different, e.g., $|\textbf{v}_1|\ll|\textbf{v}_2|$. In this scenario, the velocity of the fast particle is modified by $1+f_E/4$, while the speed of the slower particle gets enhanced to a value as large as $|\textbf{v}_2|f_E/4$. In other words, for this scenario the fast particle is only perturbed slightly while the slow particle can be perturbed significantly. We comment that more strongly perturbing slower particles is preferred as perturbations to the lower velocity part of the distribution function typically has less of an effect on the physics.

\subsection{Discussion of potential failures of the energy-correction algorithm}

The energy-correction method works by adjusting the weighted-CM energy of random binary pairs after the momentum correction has been applied. If the violation of energy conservation, $\Delta E$ is large in magnitude compared to the combined weighted-CM energy of all binary pairs, then the method can fail. We find that this can occur when there are very few particles in a cell (e.g., $N$ order 10), and at least one of the particles has a weight that is much larger than the others. When the large weight particle is determined to scatter, it can cause a relatively large violation in energy conservation in the cell that cannot be re-absorbed by adjusting the weighted-CM energy of the very few binary pairs in the cell. We have only encountered this problem with the number of particles in cell being order $10$. For such a low number of particles in a cell, one cannot expect physically accurate results for scattering with weighted particles. The simplest solution is to do nothing and live with a momentum-conserving but not energy-conserving solution. The next simplest solution is to save the momentum vectors of the particles prior to scattering and, if the method fails, reset the particles velocities and skip doing scattering in this cell during this time step.

\subsection{Computational cost of momentum-correction algorithm}

The moment-correction algorithm increases the computational cost of the overall scattering algorithm. For test T1d, but with $10{\times}$ more particles ($N=82,000$), the relative computational cost for no moment-correction, moment-correction without sorting, and moment-correction with sorting, are is 1.00, 1.23, and 1.72, respectively. The computational cost of applying the moment-correction algorithm itself increases the cost by $23$\%. The cost increases by an additional $40$\% when sorting the particles. We are currently using the standard std::sort() function, which may not always have the ideal computational cost of order $N\log\left(N\right)$. In the future, we will seek out more efficient sorting algorithms to reduce the computational cost. Furthermore, it is not always necessary to sort all of the particles (see Fig.\,\ref{T1d_count}). The computational burden of sorting can also be greatly reduced by only doing partial sorting.

\subsection{Summary}

A method for doing weighted-particle Coulomb collisions in a plasma that preserves scattering physics while exactly conserving momentum and energy is described in this manuscript. We first show how to extend $N{\times}N$ scattering methods for equally weighted particles to arbitrarily-weighted particles such that the normalized scattering length is correct for each particle on average. Then, it is straightforward to deduce an expression for the scattering length of binary pairs for order $N$ methods that is a direct generalization of that given by Takizuka and Abe \cite{Takizuka1977205} to include weighted particles. 

The weighted-particle scattering method alone only preserves momentum and energy on average. A new method is introduced to rigorously restore exact momentum and energy conservation post scatter. The method, which is based on inelastic scattering dynamics, works by forming random binary pairs and making small adjustments to the weighted CM energy of each pair. The method works for both non-relativistic and relativistic particles. The efficacy of the entire algorithm is illustrated with various test problems, including those in which the original Nanbu method for weighted-particle Coulomb scattering \cite{Nanbu1998639} is known to produce incorrect results \cite{HIGGINSON2020109450}.

While this work focused on small angle Coulomb collisions, the extension to weighted particles and the method to restore momentum and energy conservation are not specific to Coulomb collisions in any way. These methods work equally well for other collision processes too, such as those governed directly by the Boltzmann collision integral discussed at the beginning of Sec.\,\ref{Sec:Boltzmann} where the normalized path length is used as a probability of a collision occurring rather than as a variance for a Gaussian distribution function from which a scattering angle is sampled from.

\appendix
\section{Extension of moment-correction method to relativistic particles} \label{appendixA}

Restoring momentum conservation for relativistic particles is done the same as it is for non-relativistic particles, The only difference is that the proper velocity ($\textbf{u}_p=\gamma_p\textbf{v}_p$) is shifted rather than the actual velocity. The method for correcting for energy conservation, on the other hand, must be done a bit differently. For relativistic particles, a Lorentz transformation of the four-momentum vector must be done to go between the lab and center-of-momentum frames (relativistic analog of the center-of-mass frame). The CM velocity for a relativistic treatment is defined as
\begin{equation}
\textbf{v}_{\text{CM}} = \frac{\textbf{p}_1+\textbf{p}_2}{\gamma_1m_1 + \gamma_2m_2} = \frac{\textbf{p}_1+\textbf{p}_2}{E_{tot}/c^2}. \label{Eq:Vcm}
\end{equation}

One could attempt the same procedure as before; transform to the weighted-CM frame, adjust the magnitude of the particle velocities to account for an inelastic energy sink/source, then transform back to the lab frame. However, there is a subtlety with respect to inelastic processes in relativistic mechanics that needs to be mentioned. The potential energy of an inelastic process in relativistic mechanics is realized via a change in particle mass. This is well known for fusion processes, but it is the same for any inelastic process. This means that the mass of the particles change in an inelastic process such that the total energy, $E_{tot} = \left(\gamma_1m_1 + \gamma_2m_2\right)c^2$, does not change. In other words, if we transform to the CM frame, alter the kinetic energy but not the mass of the particles, and then transform back to the lab frame, we will find that 1) momentum is not conserved and 2) the change in kinetic energy does not match that in the CM frame. The reason is that the CM frame is altered since $E_{tot}$ in Eq.\,\ref{Eq:Vcm} has changed.

To perform the energy-correction method for relativistic particles we write the change in total energy and momentum as follows
\begin{eqnarray}
E'_{tot} &=& E_{tot} - U, \label{deltaE_rel} \\
\textbf{p}'_1 &=& \textbf{p}_1 + \alpha \Delta\textbf{u}, \label{p1p_rel} \\
\textbf{p}'_2 &=& \textbf{p}_2 - \alpha \Delta\textbf{u}, \label{p2p_rel}
\end{eqnarray}
where $\Delta\textbf{u}\equiv\textbf{u}_1 - \textbf{u}_2$ and $U\equiv\text{sign}\left(\Delta E\right)f_EKE_{cm}$ is defined the same as before with $KE_{cm}$ the kinetic energy in the weighted-CM frame. From Lorentz invariance of the product of 4-vectors, the total energy in the CM frame is $E^2_{cm}=E^2_{tot} - |\textbf{p}_{tot}|^2c^2$ where $\textbf{p}_{tot}=\textbf{p}_1+\textbf{p}_2$. The kinetic energy in the CM frame is $KE_{cm} = E_{cm} - \left(m_1+m_2\right)c^2$. The scalar quantity $\alpha$ in Eqs.\,\ref{p1p_rel}-\ref{p2p_rel} is computed using the following relativistic relationship between energy and momentum:
\begin{eqnarray}
\frac{E_{tot}-U}{c^2} = m_1\gamma'_1 + m_2\gamma'_2 = \sqrt{m_1^2 + |\textbf{p}'_1|^2/c^2} + \sqrt{m_2^2 + |\textbf{p}'_2|^2/c^2}. \label{alpha_Eq0}
\end{eqnarray}
After some algebraic manipulation, Eq.\,\ref{alpha_Eq0} can be written as the following quadratic equation for $\alpha$:
\begin{equation}
\frac{4}{c^4}\left[A^2|\Delta\textbf{u}|^2c^2 - \left(\textbf{p}_{tot}\cdot\Delta\textbf{u}\right)^2\right]\alpha^2 + \frac{4}{c^2}\Delta\textbf{u}\cdot\left[D\textbf{p}_{tot}-2A^2\textbf{p}_2\right]\alpha + \frac{4}{c^4}A^2E_2^2 - D^2 = 0, \label{alpha_Eq1}
\end{equation}
where $A \equiv \left(E_{tot}-U\right)/c^2$ and $D \equiv A^2 + \left(E_2^2 - E_1^2\right)/c^4$. The solution of Eq.\,\ref{alpha_Eq1} has two roots. The positive root is the smaller of the two and is the solution that converges to the same expression used for the non-relativistic treatment in the limit where the particle velocities are small compared to the speed of light.

\section*{Acknowledgments}
This work was performed under the auspices of the U.S. Department of Energy by Lawrence Livermore National Laboratory under Contract DE-AC52-07NA27344 and was supported by the LLNL-LDRD Program under Project No. 23-ERD-007.

\bibliographystyle{model1-num-names}

\end{document}